\documentclass[11pt,fleqn, reqno]{amsart}
\usepackage{latexsym}
\usepackage{amssymb}
\usepackage{amsmath}
\usepackage{amsthm}

\begin{document}
\title{Symmetries and Pre-metric Electromagnetism}

\author{\textit{David} \textit{DELPHENICH}}
\maketitle
\textit{Physics Department, Bethany College, Lindsborg, KS 67456 USA}

E-mail: \textit {delphenichd@bethanylb.edu}

Received October, 2005, in final form ???, Published online ???

Original article is available at 
{http://www.emis.de/journals/SIGMA/2005/Paper??/}

\begin{abstract}
The equations of pre-metric electromagnetism are 
summarized and then formulated as an exterior differential system on the 
total space of the bundle of 2-forms over the spacetime manifold. The 
Harrison-Estabrook method of computing the symmetries of the system is then 
applied, with the result that of the four possible formal algebras of 
infinitesimal symmetries, the most physically compelling one is the Lie 
algebra of infinitesimal projective transformations of RP$^{4}$.
\end{abstract}

\textit{Key words: Symmetries of differential equations, pre-metric electromagnetism, Lie equations, formal algebras.}

\textit{2000 Mathematics Subject Classification: 35A30, 58J70, 53C80, 78A02}

\section{Introduction}

Something that is easier to understand in terms of modern mathematical 
formalism that was not nearly as clear in the early Twentieth Century was 
the fact that one must clearly distinguish between the symmetries of a 
system of differential equations in a set of field (scalar, tensor, etc) 
variables, in the sense of transformations of the coordinates and field 
components that preserve the \textit{form} of the equations, and symmetries in the sense of transformations that take \textit{solutions} of the system of equations to other solutions. It is the latter class of transformations that we address in the present study and refer to as the symmetries of that system of partial differential equations that is defined by the pre-metric formulation of 
Maxwell's equations of electromagnetism.

It was Lorentz who first showed that the symmetries of the 
space of solutions to Maxwell's equations, in their traditional 
metric-dependent form, included the Lorentz transformations.  This result was expanded by Bateman \cite{Bateman} and Cunningham \cite{Cunningham}, who showed 
that, in fact, the finite-dimensional part of the symmetry group was the 
conformal Lorentz group, which consists of all diffeomorphisms of Minkowski 
space that preserve the light cone. 

Some years later, Harrison and Estabrook 
\cite{HandE} formulated the problem of finding the symmetries of the 
solution space to the Maxwell equations in terms of the method of exterior 
differential systems, which goes back to Cartan \cite{Cartan1} and K\"{a}hler 
\cite{Kaehler} (see also Choquet-Bruhat \cite{Choquet}). They obtained the same symmetries as Bateman and Cunningham, along with the 
infinite dimensional Abelian group that is defined by the addition of any 
other solution, which one would expect from the linearity of the system. 
Furthermore, this linearity suggests that the solutions should be invariant 
under scalar multiplication of the field components.  In addition, one has the symmetry defined by the Hodge duality isomorphism acting on the 2-forms.  Since this isomorphism defines an almost-complex structure on the vector bundle of 2-forms, one finds that the symmetry under multiplication by real scalars can be extended to multiplication by complex scalars.

More recently, a comprehensive presentation of the symmetries of Maxwell's equations was given by Fushchich and Nikitin \cite{Fushchich}.

Now, it was first observed by Cartan \cite{Cartan2}, and later expanded upon 
by Kottler \cite{Kottler} and Van Dantzig \cite{DVD}, that the only place 
where spacetime metric appears in Maxwell equations is in the Hodge duality 
isomorphisms:
\begin{displaymath}
\ast : \Lambda ^{\ast }(M) \to \Lambda ^{4-\ast 
}(M)$, \quad \textit{$\alpha $} $\mapsto $ *\textit{$\alpha $}$. 
\end{displaymath}

Kottler and Van Dantzig then succeeded in re-formulating Maxwell's equations 
without the introduction of the usual Lorentzian pseudo-metric, but by 
substituting an electromagnetic constitutive law as the agent of this new 
formulation.

Along with the aforementioned purely mathematical consideration, one also must confront the purely physical consideration that the appearance 
and nature of the spacetime pseudo-metric of relativity theory is intimately 
linked with the propagation of electromagnetic waves, even though -- 
paradoxically -- the metric structure of spacetime seems to be ultimately 
more fundamental to the presence of \textit{gravitational} forces thatn it is to the presence of electromagnetic ones. Hence, there is reason to suspect that the much weaker gravitational force 
is, in some sense, subordinate to the much stronger electromagnetic one. In 
particular, one can define the Lorentzian structure as something that 
appears by way of the principal symbol of the d'Alembertian operator, and 
can be derived from the electromagnetic constitutive law by the use of the 
Fresnel analysis of waves in anisotropic media \cite{Hehl1}, suitably 
adapted to four-dimensional methods.

One gets more geometrically fundamental isomorphisms than Hodge duality from the Poincar\'{e} duality between $\Lambda _{\ast }$($\mathbb{R} ^{4})$ and $\Lambda $*($\mathbb{R}^{4})$ that is defined by a choice of unit volume element. Indeed, Hodge duality can be obtained by composing the isomorphisms of Poincar\'{e} duality with the isomorphisms of k-vectors and k-forms that one obtains from the metric-defined isomorphism of tangent spaces with cotangent spaces.  Since Poincar\'{e} duality is best presented in the framework of projective geometry, this suggests that projective geometry might be more appropriate for the treatment of electromagnetism than metric geometry.

Moreover, the conformal Lorentz group is associated with the introduction of 
light cones into the tangent spaces of the spacetime manifold. Indeed, 
physically, the measurement of distances in spacetime is facilitated by the 
introduction of electromagnetic waves. However ``most'' bivectors (2-forms, 
resp.) are not wavelike, so the use of a structure -- namely, the Lorentzian 
structure -- that is associated with the wave solutions 
restricts the generality of Maxwell's equations. Hence, we must treat 
electromagnetism as something that is broader in geometrical scope than the 
introduction of even a conformal structure might suggest.

Finally, a compelling indication of the propriety of projective geometry 
comes from the consideration of the symmetries of the pre-metric Maxwell 
equations, in the sense of the symmetries of their space of solutions. In 
the present study\footnote{ The essential results of this presentation were 
presented in more detail in a previous work by the author \cite{DHD1}.\par 
}, the Harrison-Estabrook method for finding symmetries of systems of 
differential equations is applied to the case of the equations of pre-metric 
electromagnetism. It is found that although, in the absence of deeper 
analysis, there seems to be a choice of four possible symmetry groups for 
PMEM\footnote{ For the sake of brevity, in the sequel, we shall refer to the 
resulting theory of ``pre-metric electromagnetism'' by the acronym PMEM.}, 
nevertheless, the one that seems to most directly extend the conformal 
Lorentz symmetry that was established by Bateman and Cunningham is the group 
\textit{SL}(5; $\mathbb{R} $), which represents the group of projective transformations of $\mathbb{R} $P$^{4}$.

\section{Pre-metric Maxwell equations \cite{Hehl1, DHD1, DHD2, Post}}

We assume that our spacetime manifold $M $is four-dimensional, orientable, and 
given a specific choice \textit{$\varepsilon $} $\in  \Lambda ^{4}(M)$ of unit-volume element on 
$T(M)$. One can then define a unit-volume element \textbf{$\epsilon $} $\in $ 
$\Lambda _{4}(M)$ on $T$*($M)$ by choosing the unique 4-vector field 
\textbf{$\epsilon $} such that \textit{$\varepsilon $}(\textbf{$\epsilon $}) = 1. For a natural frame field $\partial _{\mu }=\partial $/$\partial x^{\mu }$ 
that is defined by a local coordinate chart ($U$, $x^{\mu })$ on an open subset 
$U$ in $M$, and its reciprocal local co-frame field \textit{dx}$^{\mu }$, these two volume 
elements take the local forms:
\begin{displaymath}
$\textit{$\varepsilon $} = \textit{dx}$^{0} \wedge $ \textit{dx}$^{1} \wedge $ \textit{dx}$^{2} \wedge $ \textit{dx}$^{3}=\frac{1}{4!}$\textit{$\varepsilon $}$_{\kappa 
\lambda \mu \nu }$\textit{ dx}$^{\kappa } \wedge $ \textit{dx}$^{\lambda } \wedge $ \textit{dx}$^{\mu } \wedge $
\textit{dx}$^{\nu },
\end{displaymath}
\begin{displaymath}
$\textbf{$\epsilon $} = $\partial _{0} \wedge \partial _{1} \wedge 
\partial _{2} \wedge \partial _{3}=\frac{1}{4!}$\textit{$\epsilon $}$^{\kappa \lambda \mu \nu } \partial _{\kappa } \wedge \partial _{\lambda } 
\wedge \partial _{\mu } \wedge \partial _{\nu }
\end{displaymath}

A first key point of departure of PMEM from the conventional formulation of 
Maxwell's equations is the fact the divergence operator on $\Lambda _{\ast 
}(M)$, viz., \textit{$\delta $} $\equiv $ {\#}$^{-1}d${\#}, is defined by the \textit{Poincar\'{e} duality} isomorphism:
\begin{displaymath}
${\#}: $\Lambda _{\ast }(M) \quad \to \Lambda ^{4-\ast }(M)$, \textbf{a} $\mapsto  \quad i_{a}$ \textit{$\varepsilon $}= \textit{a}$ ^{\mu \ldots \nu }$ \textit{$\varepsilon $}$_{\kappa \lambda \mu \nu }.
\end{displaymath}

\noindent which comes from the volume element, not Hodge duality, which requires a metric, in addition. Indeed, this is an important subtlety concerning the 
divergence operator in general, which is often presented as something that 
requires the introduction of a metric for its definition.

A second point of departure is that the role of an explicitly specified 
electromagnetic constitutive law is given more prominence than in most 
treatments of Maxwell's equations using exterior differential forms. In 
general, an \textit{electromagnetic constitutive law} takes the form of an invertible fiber-preserving map:
\begin{displaymath}
$\textit{$\chi $}: $\Lambda ^{2}(M) \to  \Lambda _{2}(M)$, $F \mapsto  \mathfrak{h}$ = \textit{$\chi $}($F),
\end{displaymath}
\noindent that is a diffeomorphism of the fibers in the nonlinear case and a linear isomorphism:
\begin{displaymath}
\mathfrak{h}^{\mu \nu }=\tfrac{1}{2}F_{\kappa \lambda }$
\textit{$\chi $}$^{\kappa \lambda \mu \nu }
\end{displaymath}

\noindent in the linear case. The bivector field $\mathfrak{h}$ that corresponds to a given 2-form $F $is referred to as its \textit{electromagnetic excitation} bivector field.

If $F$ is the usual Minkowski 2-form of electromagnetic field strengths, $d$ is 
the exterior derivative operator on $\Lambda $*($M)$, and \textbf{J} is the 
vector field of electric current (the source of the electromagnetic field) 
then the pre-metric Maxwell equations take the form:
\begin{equation}
$\textit{dF }= 0, \quad \textit{$\delta $} $\mathfrak{h}$ = \textbf{J}, 
\quad $\mathfrak{h}$ = \textit{$\chi $}$(F).
\end{equation}

In local form, these are:
\begin{displaymath}
\partial F_{\lambda \mu \nu }+\partial F_{\mu \nu \lambda }
+\partial F_{\nu \lambda \mu }$ = 0, \quad
$\partial _{\mu } F^{\mu \nu }=J^{\nu }$, \quad
$\mathfrak{h}^{\mu \nu }$ = 
\textit{$\chi $}$(F_{\mu \nu }).
\end{displaymath}

We can compose the isomorphisms \textit{$\chi $} and {\#} to obtain another isomorphism \textit{$\kappa $ }= {\#} $\cdot $\textit{$\chi $} of the bundle $\Lambda ^{2}(M)$ with itself. By assumption, 
the isomorphism \textit{$\kappa $} is proportional to an almost-complex structure on \textit{$\kappa $}$^{2}$ = 
-- \textit{$\lambda $I}, so it is equivalent, up to a multiplicative scalar function, to the isomorphism that one would have obtained from the Hodge *, had one introduced a Lorentzian structure on $T(M)$. We can also treat \textit{$\kappa $} as fourth rank tensor field on $M$ whose component form is:
\begin{displaymath}
$\textit{$\kappa $} =$\frac{1}{4}$\textit{$\kappa $}$_{\mu \nu }^{\alpha \beta }$ ($\theta ^{\mu } \wedge 
\theta ^{\nu }) \otimes $ (\textbf{e}$_{\alpha } \wedge $ 
\textbf{e}$_{\beta })$ .$
\end{displaymath}

\section{Electromagnetic constitutive laws \cite{Hehl1}, \cite{DHD1}, \cite{DHD2}, \cite{Post}, \cite{LLP}, \cite{Hehl2}, \cite{Hehl3}, \cite{Lindell}}

Clearly, the pre-metric Maxwell equations (1) closely resemble the usual 
Maxwell equations, except that the role of the electromagnetic constitutive 
law has supplanted that of the Lorentzian pseudo-metric. Hence, we now 
briefly discuss both the physical and mathematical aspects of postulating an 
electromagnetic constitutive law as a fundamental object.

In classical vacuum (Maxwellian) electromagnetism, $\mathfrak{h}$ is linear on the fibers, and if the Minkowski 2-form takes the local form:
\begin{displaymath}
F =\tfrac{1}{2}F_{\mu \nu }$\textit{ dx}$^{\mu } \wedge $ \textit{dx}$^{\nu }=E_{i}$ 
\textit{dx}$^{0} \wedge $ \textit{dx}$^{i}+\tfrac{1}{2}$\textit{$\varepsilon $}$_{ijk }B^{i}$ \textit{dx}$^{j} \wedge $ 
\textit{dx}$^{k}
\end{displaymath}

\noindent then \textit{$\chi $ }takes the linear homogeneous isotropic form:
\begin{displaymath}
$\textit{$\chi $}($F)=D^{i} \partial _{0} \wedge \partial _{i}$ 
+$\tfrac{1}{2}$\textit{$\varepsilon $}$^{ijk }H_{i} \partial _{j} \wedge \partial _{k}$ = 
\textit{$\varepsilon $}$_{0} \delta ^{ij} E_{j} \partial _{0} \wedge 
\partial _{i}$ + 
$\frac{1}{2\mu _0 }$\textit{$\varepsilon $}$^{ijk }B_{i} \partial _{j} \wedge \partial _{k}$ .$
\end{displaymath}

\noindent in which the indices $i, j, k$ range over the spatial values 1, 2, 3.

The constant \textit{$\varepsilon $}$_{0}$ is referred to as the \textit{electric permeability} (or \textit{dielectric constant}) of the vacuum and \textit{$\mu $}$_{0}$ is 
its \textit{magnetic permeability.}

In both of the expressions for $F$ and \textit{$\chi $}($F)$ we have implicitly used the Euclidian 
spatial metric, whose components in the chosen frame are \textit{$\delta $}$_{ij}$, and its 
inverse \textit{$\delta $}$^{ij}$, to raise and lower the index of $B_{i}$ and $H^{i}$, 
respectively. Hence, one should carefully note that the expression for 
\textit{$\chi $}($F)$ is not actually invariant under Lorentz transformations of the local frame field $\partial _{\mu }$, but only under Euclidian 
rotation of its spatial members. 

What makes this intriguing is that the speed of propagation for electromagnetic waves in vacuo is \textit{derived} from \textit{$\varepsilon $}$_{0}$ and \textit{$\mu $}$_{0}$:
\begin{displaymath}
c_{0}=\frac{1}{\sqrt {\varepsilon _0 \mu _0 } }$.$
\end{displaymath}

\noindent Hence, a fundamental assumption of special relativity -- viz., that 
$c_{0}$ is a constant that is independent of the choice of Lorentz frame -- 
seems to resolve at the pre-metric level to the statement that the constants 
from which $c_{0}$ is constructed can change with the choice of Lorentz 
frame, but must do so in a complementary way.

One can eliminate the homogeneity restriction and allow \textit{$\varepsilon $} and \textit{$\mu $} to vary with 
position, which is essentially what one does in linear optics. In such a 
case, it is usually not the speed of propagation in the medium that one 
considers, but its index of refraction:
\begin{displaymath}
n(x)=\frac{c_0 }{c(x)}=\sqrt {\frac{\varepsilon _0 \mu _0 }{\varepsilon 
(x)\mu (x)}} $.$
\end{displaymath}

Furthermore, one can drop the isotropy restriction, in which 
case,\textit{$\varepsilon $}$_{0}$ and \textit{$\mu $}$_{0}$ are replaced by 3$\times $3 matrices whose components 
are functions of position and possibly time. This situation relates to the 
propagation of electromagnetic waves in crystal media, in which the speed of 
propagation can vary with direction, as well as position.

The case of a nonlinear \textit{$\chi $} not only has immediate application to nonlinear 
optics, but also a possible application to effective QED, perhaps in the 
effective models for vacuum polarization, such as the Born-Infeld model.

Returning to the linear case, \textit{$\chi $} also defines a non-degenerate bilinear form 
on $\Lambda ^{2}(M)$:
\begin{displaymath}
$\textit{$\chi $}($F$, $G) \equiv  G$(\textit{$\chi $}($F))$ = 
\textit{$\chi $}$_{IJ}F^{I}G^{J}=
\tfrac{1}{2}$
\textit{$\chi $}$_{\kappa \lambda \mu \nu } F^{\kappa \lambda } G^{\mu \nu }
\end{displaymath} 

This bilinear form admits a decomposition that is irreducible under the 
action of \textit{GL}(6; $\mathbb{R}$) on $\Lambda ^{2}(M) \otimes \Lambda ^{2}(M)$ by congruence:
\begin{displaymath}
$\textit{$\chi $} = 
$^{(1)}$ \textit{$\chi $} + 
$^{(2)}$ \textit{$\chi $} + 
$^{(3)}$ \textit{$\chi $}$
\end{displaymath}

\noindent in which:
\begin{displaymath}
^{(1)}$ \textit{$\chi $}= 
 \textit{$\chi $} -- $^{(2)}$ \textit{$\chi $} –- $^{(3)}$ \textit{$\chi $}$
\end{displaymath}

\noindent is called the \textit{principal part}. It is symmetric and ``traceless,'' in the sense that it does 
not contain a contribution that is proportional to the volume element 
\textit{$\varepsilon $}.

The tensor field:
\begin{displaymath}
^{(2)}$\textit{$\chi $} = $\tfrac{1}{2}$(\textit{$\chi $} -- \textit{$\chi $}$^{T})
\end{displaymath}

\noindent is the anti-symmetric \textit{skewon} part of \textit{$\chi $}. It is associated with established physical 
phenomena, such as the Faraday effect, and natural optical activity 
\cite{LLP}, \cite{Hehl2}, \cite{Hehl3}.

The tensor field:
\begin{displaymath}
^{(3)}$\textit{$\chi $}=\textit{$\chi $}$(E_{I}$, $E_{I})$ 
\textit{$\varepsilon $}$
\end{displaymath}

\noindent is the \textit{axion} part of \textit{$\chi $}, which is proportional to the volume element. It does not affect 
the propagation of electromagnetic waves, but Lindell and Sivola 
\cite{Lindell} have suggested that it might still play a role in some 
electromagnetic media.

In the language of projective geometry, the case of a general, but linear, 
\textit{$\chi $} defines a \textit{correlation} on the fibers of $\Lambda ^{2}(M)$ (more precisely, their 
projectivizations), namely, a linear isomorphism from each fiber of $\Lambda 
^{2}(M)$ to its dual vector space, which is a fiber of $\Lambda 
_{2}(M)$. A symmetric correlation is called a \textit{polarity} and defines a quadratic 
form. An anti-symmetric correlation defines a \textit{null polarity}, much like a symplectic form 
on an even-dimensional vector space.

The manner by which \textit{$\chi $} gives rise to a Lorentzian metric on $T(M)$ follows from 
adding certain restricting assumptions on \textit{$\chi $}. Essentially, one looks for an 
``exterior square root'' \textit{$\chi $} = ``$g \wedge g$,'' since, in the case of the Hodge * 
isomorphism, the role of \textit{$\chi $} is played by the map \textit{$\iota $}$_{g} \wedge $ \textit{$\iota $}$_{g}$: $\Lambda 
^{2}(M)\to \Lambda _{2}(M)$ whose tensor components are:
\begin{displaymath}
\tfrac{1}{2}$(g$^{\kappa \lambda }$g$^{\mu \nu }$ -- g$^{\kappa \nu }$ 
g$^{\mu \lambda }$ ) .$
\end{displaymath}

Physically, the absence of birefringence is often a necessary and sufficient 
restricting assumption for the latter reduction to take place. (Birefringence 
is an optical phenomenon that takes the form of the speed of light in a 
medium -- hence, the index and angle of refraction -- depending on the 
polarization direction of the light wave \cite{LLP}.)

\section{PMEM as an exterior differential system}

The concept of an \textit{exterior differential system} \cite{Cartan1}, \cite{Kaehler}, \cite{Choquet} is a generalization of the concept of a 
system of differential equations on a differentiable manifold. Perhaps the 
most general way of characterizing an exterior differential system is that 
it is an ideal $\mathcal{I}$ in the exterior algebra $\Lambda $*($M)$ of a manifold $M$, whose 
generators {\{}$\Theta ^{1}$, {\ldots}, $\Theta ^{k}${\}} each define a 
sub-bundle of $T(M)$ by their annihilating subspaces, which will be spaces of tangent vectors that satisfy the exterior equations:
\begin{displaymath}
\Theta ^{i}$ = 0, $i $= 1, {\ldots}, $k$ .$
\end{displaymath}
\noindent The elements of $\mathcal{I}$  then take the form \textit{$\alpha $}$_{j} \wedge \Theta ^{j}$ for suitable forms \textit{$\alpha $}$_{j }$.

The intersection of the annihilating subspaces for  all $i$  at each point defines an \textit{integral element} of the exterior differential system defined by $\mathcal{I}$; the integral submanifolds of the resulting bundle of integral elements define a foliation of $M$.

The integrability condition for such an exterior differential system is 
given by Frobenius as either $d \mathcal{I} \subset \mathcal{I}$, or, in terms of the generators of $\mathcal{I}$:
\begin{displaymath}
\Theta ^{i} \wedge d\Theta ^{i}$ = 0 for all $i$.$
\end{displaymath}

The manifold on which one needs to define the exterior differential system 
for PMEM is $\Lambda ^{2}(M)$. The equations take the local form:
\begin{displaymath}
0 = \Theta ^{1}$ = \textit{$\delta $}$^{\alpha }_{\mu }$\textit{$\delta $} $^{\beta }_{\nu }$\textit{ dF}$_{\alpha \beta } \wedge $
\textit{dx}$^{\mu } \wedge $ \textit{dx}$^{\nu }$,$
\end{displaymath}
\begin{displaymath}
0 = \Theta ^{2}$ = 
\textit{K}$_{\mu \nu }^{\alpha \beta }$ 
\textit{dF}$_{\alpha \beta } \wedge $ 
\textit{dx}$^{\mu } \wedge $ 
\textit{dx}$^{\nu }$ +
\textit{F}$_{\alpha \beta }$ 
\textit{$\kappa $}$_{\mu \nu }^{\alpha \beta }$,$_{\lambda }$
\textit{dx}$^{\lambda } \wedge $
\textit{dx}$^{\mu } \wedge $
\textit{dx}$^{\nu }$ ,$
\end{displaymath}

\noindent where:
\begin{displaymath}
$\textit{K}$_{\mu \nu }^{\alpha \beta }$ = 
\textit{$\kappa $}$_{\mu \nu }^{\alpha \beta }$ + 
$F_{\alpha \beta } \kappa _{\mu \nu }^{\alpha \beta}$ $^{,\kappa \lambda } 
\end{displaymath}

\noindent is the deformed constitutive tensor.

The expressions $F_{\alpha \beta }$, $x^{\mu }$, \textit{$\kappa $}$_{\mu \nu }^{\alpha \beta }$ are smooth functions on $\Lambda ^{2}(U)$, now, not 
smooth functions on $M$.

In general, the components of \textit{$\kappa $ } will be functions of both position and field strength, so:
\begin{displaymath}
$\textit{d$\kappa $}$_{\mu \nu }^{\alpha \beta }$ = 
\textit{$\kappa $}$ _{\mu \nu }$,$_{\lambda }^{\alpha \beta }$
\textit{dx} $^{\lambda }$ + 
\textit{$\kappa $}$_{\mu \nu }^{\alpha \beta }$ $^{,\kappa \lambda }$
\textit{dF}$_{\kappa \lambda }
\end{displaymath}

This expression defines three special cases of \textit{$\kappa $}$_{\mu \nu }^{\alpha \beta }$

$i)$ The \textit{uniform linear }case:
\begin{equation}
$\textit{d$\kappa $}$_{\mu \nu }^{\alpha \beta } = 0.
\end{equation}

This case essentially describes the classical vacuum, for which the 
independent non-zero components of 
\textit{$\kappa $}$_{\mu \nu }^{\alpha \beta }$ 
are the vacuum permittivity \textit{$\varepsilon $}$_{0}$ and the inverse vacuum magnetic permeability \textit{$\mu $}$_{0}$ , both of which are assumed to be constant, although, as pointed 
out above, this constancy is not frame-invariant.

\textit{ii}) The \textit{non-uniform linear} case:
\begin{equation}
$\textit{$\kappa $}$_{\mu \nu }$,$_{\lambda }^{\alpha \beta } \ne $ 0, \quad
\textit{$\kappa $}$_{\mu \nu }^{\alpha \beta }$ $^{,\kappa \lambda } = 0 .
\end{equation}

This case is most appropriate to linear optical media, such as linear 
isotropic media, in which now the functions \textit{$\varepsilon $}($x)$ and \textit{$\mu $}($x)$ are assumed to vary 
with position, but not field strength. This has the consequence that speed 
of light -- or its inverse, the index of refraction -- might vary from point 
to point, as well.

\textit{iii}) The \textit{uniform nonlinear }case:
\begin{equation}
$\textit{$\kappa $}$_{\mu \nu }$,$_{\lambda }^{\alpha \beta }$ = 0, \quad
\textit{$\kappa $}$_{\mu \nu }^{\alpha \beta }$ $^{,\kappa \lambda } \ne 0.
\end{equation}

This case might describe either nonlinear optical media, in which the 
nonlinearity is related solely to the field strengths, but not the position 
in the medium, or even possibly the quantum vacuum, which seems to be 
characterized by the onset of vacuum polarization at field strengths beyond 
some critical limit. One might then investigate uniform nonlinear isotropic 
media, for which the constants would take the forms \textit{$\varepsilon $}($F)$ and \textit{$\mu $}($F)$. Then again, 
if the polarization of a dielectric medium is analogous to the magnetization 
of a magnetic medium, one might have to consider the spontaneous breaking of 
the vacuum symmetry under polarization, which would suggest anistropy.

\section{Symmetries of exterior differential systems \cite{HandE}, \cite{Olver}, \cite{Bluman} }

A \textit{finite symmetry} of an exterior differential system $\mathcal{I}$ on a manifold $M$ is a diffeomorphism $f$ of $M $to itself that takes solutions -- i.e., integral submanifolds -- of $\mathcal{I}$ to other solutions.  As a result:
\begin{displaymath}
f$*$\mathcal{I} = \mathcal{I}$.$
\end{displaymath}

Hence, if $\mathcal{I}$ is generated by the set {\{}$\Theta ^{i}$, $i $= 1, {\ldots}, $p$ {\}} then:
\begin{displaymath}
f$*$\Theta ^{i}=\alpha _j^i \wedge \Theta ^{j}
\end{displaymath}

\noindent for forms $\alpha _j^i $ of suitable degree.

An \textit{infinitesimal symmetry} of $\mathcal{I}$ is a vector field $X$ on $M$ that satisfies a \textit{Lie equation} \cite{KandS} of the form:
\begin{displaymath}
$L$_{X} \mathcal{I} \subset \mathcal{I}$.$
\end{displaymath}

\noindent Hence:
\begin{displaymath}
$L$_{X}\Theta ^{i}=\alpha _j^i \wedge \Theta ^{j}
\end{displaymath}

\noindent for some other forms, which, by abuse of notation, we also denote by $\alpha _j^i $.

In the case of PMEM this condition takes the form:
\begin{displaymath}
$L$_{X} \Theta ^{i}$ = 
\textit{$\alpha $}$^{i} \Theta ^{1}$ + 
\textit{$\beta $}$^{i} \Theta ^{2}
\end{displaymath}
\begin{equation}
$= [\textit{$\alpha $}$^{i} \delta ^{\alpha }_{\mu }$
\textit{$\delta $}$^{\beta }_{\nu }$ +
\textit{$\beta  $}$^{ i} K_{\mu \nu }^{\alpha \beta }$] 
\textit{dF}$_{\alpha \beta } \wedge $ \textit{dx}$^{\mu } \wedge $
\textit{dx}$^{\mu }$ + 
\textit{$\beta  $}$^{ i} F_{\alpha \beta }$
\textit{$\kappa $}$_{\mu \nu }^{\alpha \beta }$ $_{,\lambda }$ 
\textit{dx} $^{\lambda } \wedge $ \textit{dx} $^{\mu } \wedge $ \textit{dx}$^{\mu }
\end{equation}

\noindent in which:
\begin{displaymath}
X=X^{\mu }\frac{\partial }{\partial x^\mu } + X_{\mu \nu }$ 
$\frac{\partial }{\partial F_{\mu \nu } }
\end{displaymath}

\noindent is a general vector field on $\Lambda ^{2}(M)$, when expressed in local form over $\Lambda ^{2}(U)=U \times {\rm A}^{2}(\mathbb{R} ^{4})$.

A key role is played in these symmetry equations by vector fields $X$ such 
that:
\begin{equation}
$L$_{X} K_{\mu \nu }^{\alpha \beta }$ = 
\textit{$\alpha $} $\delta ^{\alpha }_{\mu }$
\textit{$\delta $}$^{\beta }_{\nu }$ +
\textit{$\beta  $} $K_{\mu \nu }^{\alpha \beta }
\end{equation}

\noindent for suitable functions \textit{$\alpha $}, \textit{$\beta $}. This 
equation is essentially a generalization of the condition for conformal 
Killing vector fields on a Riemannian (or pseudo-Riemannian) manifold, which 
are then the infinitesimal generators of infinitesimal conformal 
transformations for the metric in question. The vector fields that satisfy 
(6) include the ones that preserve \textit{$\kappa $}, up to a scalar multiple.

\section{Prolongations and formal algebras}

Finding the full group of symmetries of an exterior differential system 
becomes a matter of finding all of the solutions to the Lie equations that 
one obtains from the symmetry condition. For the Lie equations (5) of PMEM 
this generally means looking at increasingly more complicated forms for the 
functions \textit{$\alpha $}$^{i}$, \textit{$\beta $}$^{i}$, as one might do by expanding them into a Taylor series.

When the functions \textit{$\alpha $}$^{ i}$, \textit{$\beta $}$^{ i}$ are constant, the set of all $X$ that solve 
(5) generally defines a finite-dimensional Lie subalgebra $\mathfrak{g}$ of the Lie algebra $\mathfrak{X}(M)$ of vector fields on $\Lambda ^{2}(U)$.

When the \textit{$\alpha $}$^{ i}$ and \textit{$\beta $}$^{ i}$ are non-constant, one might express them in a 
formal\footnote{ The use of term ``formal'' is to suggest that one is not 
concerned with the convergence of the series, since one is generally only 
looking at some finite sub-series of the full infinite series.} power 
series, along with $X$ itself, and use the Lie equation to relate the 
coefficients of the two series. This will then define an infinite sequence 
of increasingly higher-order systems of PDE's.

The vector fields that satisfy each successive system of PDE's define a 
sequence of \textit{prolongations} $\mathfrak{g}^{(i)}$ of $\mathfrak{g}$ \cite{Sternberg}, \cite{Singer}, which do not themselves have 
to be Lie algebras, since they might not be closed under Lie bracket. If any 
one of them vanishes then all of the following prolongations vanish. The 
sequence of prolongations might terminate in this way after a finite number 
of steps, in which case one says that $\mathfrak{g}$ has \textit{finite type;} otherwise, it has \textit{infinite type.} For instance, $\mathfrak{so}(n; \mathbb{R})$ is of type 1, since its first prolongation vanishes, but 
$\mathfrak{sl}(n; \mathbb{R})$ -- hence, $\mathfrak{gl}(n; \mathbb{R})$ -- are of infinite type.

Of particular interest to electromagnetism is the fact that $\mathfrak{co}(p, q)$, the linear conformal Lie algebra for any scalar product of signature type ($p$, $q)$, has a non-vanishing first prolongation, but a vanishing second 
prolongation. For the case of the linear conformal Lorentz Lie algebra in 
four dimensions, the first prolongation space is four-dimensional and is 
spanned by the generators of the infinitesimal inversions, which are 
nonlinear transformations in four-dimensions, but can be linearly 
represented in hexaspherical coordinates in six dimensions. This is a 
particularly important, but tractable, example of a finite-dimensional 
subalgebra of $\mathfrak{X}(M)$ that does not generate only linear transformations.

If the Lie algebra $\mathfrak{g}$ acts on the vector space $V $then the elements of a given prolongation $\mathfrak{g}^{(i)}$, like the partial derivatives of $X$ of order $i$ + 1, are symmetric multilinear functions of the form:
\begin{displaymath}
V \times $ {\ldots} $\times  V \to $ g ($i $+ 1 factors of $V)$.$
\end{displaymath}

Starting with the Lie algebra $\mathfrak{g}$, one can form the direct sum of the sequence of prolongations:
\begin{displaymath}
\mathfrak{g} \oplus \mathfrak{g} ^{(1)} \oplus \mathfrak{g}^{(2)}$ {\ldots}$
\end{displaymath}

One defines a graded Lie algebra over this vector space by way of Lie bracket. One calls such an algebra that is defined by the space of all formal Taylor 
series of vector fields on a given space a \textit{formal algebra}.

Note that one can also prolong a given finite sequence $\mathfrak{g}_{1}$, $\mathfrak{g}_{2}$, ${\ldots}$, $\mathfrak{g}_{m}$ of Lie algebras, 
which do not by themselves have to represent a sequence of prolongations:
\begin{displaymath}
\mathfrak{g}_{1} \oplus \mathfrak{g}_{2} \oplus {\ldots} \oplus \mathfrak{g}_{m} \oplus \mathfrak{g}_m^{(1)} \oplus \mathfrak{g}_m^{(2)} \oplus {\ldots} 
\end{displaymath}

One is essentially defining the first $n $terms of a Taylor series in such a 
case and then differentiating to obtain the higher-order terms. However, the 
prolongation of such a given sequence of Lie algebras to a formal algebra 
does not have to be unique.

\section{Uniform linear case}

When we substitute an electromagnetic constitutive law of the form (2) the 
Lie equations (5) give us that the components of $X$ must have the form 
$X^{\mu } = X^{\mu }(x)$, $X_{\mu \nu }=X_{\mu \nu }(F)$ and 
satisfy the following system of PDE's:
\begin{equation}
\frac{\partial X^\alpha }{\partial x^\mu }= $ 
\textit{$\alpha $} $\delta ^{\alpha }_{\mu }$ +
\textit{$\beta  $}$^{\alpha }_{\mu } \quad $ 
(\textit{$\beta $}$^{\alpha }_{\mu } \equiv $
\textit{$\alpha $}$^{\nu }_{\beta }$
\textit{$\kappa $}$_{\mu \nu }^{\alpha \beta }$,
\textit{$\alpha $}$^{\nu }_{\nu} \ne 0)
\end{equation}
\begin{equation}
\frac{\partial X_{\mu \nu } }{\partial F_{\alpha \beta } } = $
\textit{$\alpha $}$_{1}$
\textit{$\delta $}$^{\alpha }_{\mu }$
\textit{$\delta $}$^{\beta  }_{\nu }$ +
\textit{$\beta  $}$_{1}$
\textit{$\kappa $}$_{\mu \nu }^{\alpha \beta }
\end{equation}
\begin{equation}
$L$_{X}$\textit{$\kappa $}$_{\mu \nu }^{\alpha \beta }$ =
\textit{$\alpha $}$_{2}$
\textit{$\delta $}$^{\alpha }_{\mu }$
\textit{$\delta $}$^{\beta }_{\nu }$ +
\textit{$\beta $}$_{2}$
\textit{$\kappa $}$_{\mu \nu }^{\alpha \beta }.
\end{equation}

For constant \textit{$\alpha $}$_{i}$, \textit{$\beta $}, \textit{$\beta $}$^{\alpha }_{\mu }$ these equations can be integrated immediately and we obtain vector fields of the form:
\begin{equation}
X^{\mu }$ = \textit{$\varepsilon $}$^{\mu }$ + 
\textit{$\alpha $ x}$^{\mu }$ + 
\textit{$\beta  $}$^{\alpha }_{\mu } \, x^{\mu }
\end{equation}
\begin{equation}
X^{\mu \nu }$ = \textit{$\phi $}$^{\mu \nu }$ + 
\textit{$\gamma $ F}$^{\mu \nu }$ + 
\textit{$\delta \, \kappa $}$_{\mu \nu }^{\alpha \beta } \, F^{\mu \nu }
\end{equation}

They represent infinitesimal transformations of the following types:

$i)$ Spacetime translations, dilatations, and linear transformations, which 
collectively generate the infinitesimal affine Lie algebra. The formal 
algebra that they define is then:
\begin{displaymath}
\mathbb{R}^{4} \oplus  \mathfrak{gl}(4; \mathbb{R})$ .$
\end{displaymath}

\textit{ii}) Fiber translations, dilations, and duality transformations. The formal 
algebra that they define is isomorphic to $\mathbb{C}^{3}$.

Now, as observed above, any linear system of PDE's must include fiber 
translations and dilatations by the principle of superposition. The duality 
transformation on the space of 2-forms over a four-dimensional vector space 
defines a complex structure on it, for which the * operator behaves like 
multiplication by the imaginary $i$. Hence, we suspect that the deepest 
foundations of electromagnetism are probably expressed in terms of complex 
geometry and analysis, and not just two-dimensional potential 
theory\footnote{ See also the authors more detailed discussion on the role of complex geometry in PMEM in \cite{DHD3}.}.

In order to make contact with the established results on the symmetries of 
Maxwell's equations, we must reduce the scope of the electromagnetic 
constitutive law \textit{$\kappa $} to a symmetric uniform linear one. Hence, \textit{$\kappa $} = 
$^{(1)}$\textit{$\kappa $}, so, in particular, we have assumed that the skewon and axion parts 
vanish. One then finds that \textit{$\beta $}$^{\alpha }_{\mu }$ is an infinitesimal Lorentz 
transformation, and when \textit{$\alpha $}, \textit{$\beta $}$^{\alpha }_{\mu }$ are constant the vector 
field (10) becomes the infinitesimal generator of a linear conformal Lorentz 
transformation.

When \textit{$\alpha $}, \textit{$\beta $}$^{\alpha }_{\mu }$ are not constant, it turns out that only the linear functions produce solutions to the symmetry equations, due to the fact that the second prolongation of the linear conformal Lie algebra is 0. The first prolongation of the linear conformal Lorentz Lie algebra, namely:
\begin{displaymath}
\mathbb{R}^{4} \oplus  \mathfrak{co}(3, 1) \oplus \mathfrak{co}(3, 1)^{(1)}$ .$
\end{displaymath}

\noindent is the Lie algebra of all infinitesimal conformal Lorentz transformations.  Hence, our formalism produces the same results as were previously obtained for the conventional metric form of the Maxwell equations.

When \textit{$\kappa $} is more general than its principal part -- i.e., when it is asymmetric -- there are four possible formal algebras of spacetime 
transformations that have the same starting series $\mathbb{R}^{4} \oplus  \mathfrak{co}(4)$. Two of them are of finite type, and two of them are of infinite type.

In the finite-dimensional cases, one has:

$i)$ Infinitesimal affine transformations (all prolongations are zero). This 
is simply the formal algebra:
\begin{displaymath}
\mathbb{R}^{4} \oplus \mathfrak{gl}(4)$.$
\end{displaymath}

\textit{ii}) Infinitesimal projective transformations of $\mathbb{R}$P$^{4}$. Now, the formal algebra is:
\begin{displaymath}
\mathbb{R}^{4} \oplus \mathfrak{gl}(4) \oplus \mathfrak{p}^{(1)}$ .$
\end{displaymath}

The vector space $\mathfrak{p}^{(1)}$ is an invariant four-dimensional subspace of $\mathfrak{gl}(4)^{(1)}$ that consists of the infinitesimal inversions that extend the four-dimensional affine Lie algebra to the four-dimensional projective Lie algebra.

In the infinite-dimensional cases, one has:
\textit{iii}) All vector fields, whose formal algebra is the most general one that is possible:
\begin{displaymath}
\mathbb{R}^{4} \oplus \mathfrak{gl}(4) \oplus \mathfrak{gl}(4)^{(1)} \oplus {\ldots} 
\end{displaymath}

\textit{ii}) Constant-divergence vector fields, whose formal algebra is:
\begin{displaymath}
\mathbb{R}^{4} \oplus \mathfrak{gl}(4) \oplus \mathfrak{sl}(4)^{(1)} \oplus  {\ldots}
\end{displaymath}

These vector fields are the infinitesimal generators of transformations that 
make the volume of a given region of $\mathbb{R}^{4}$ vary with time as \textit{t}$^4$, since the divergence of any vector field \textbf{v }is the Poincar\'{e} dual of the Lie derivative of the volume element in the direction \textbf{v}:
\begin{displaymath}
$L$_{\mathbf{v}}$\textit{$\varepsilon $} = \textit{di}$_{\mathbf{v}}$\textit{$\varepsilon $} = $d${\#}\textbf{v} = {\#}\textit{$\delta $}\textbf{v}.$
\end{displaymath}

\noindent Hence, if \textit{$\delta $}\textbf{v} = $\lambda $ then:
\begin{displaymath}
$L$_{\mathbf{v}}$\textit{$\varepsilon $} = $\lambda $ \textit{$\varepsilon $}.$
\end{displaymath}

Of course, there is something physically unsatisfying about the idea that 
the symmetries of the space of solutions of the pre-metric Maxwell equations 
seem to present physics with a decision to make between four possibilities. 
Clearly, the rigorous basis for making that choice demands further analysis 
of the role that the symmetries play in physical phenomena and how they 
relate to the structure of \textit{$\kappa $}. However, since one would expect that making \textit{$\kappa $} more general than the metric-dependent expression in Maxwell's equations would suggest an expansion of scope in the symmetries, that would seem to rule out the affine Lie algebra as a likely possibility. The possibility that most closely extends the conformal Lorentz Lie algebra of the metric form of Maxwell's equations is the Lie algebra of infinitesimal projective transformations of $\mathbb{R}$P$^{4}$. The formal algebra of constant divergence vector fields is what one would expect in the absence of \textit{$\kappa $}, but the presence of the volume element \textit{$\varepsilon $}, and the formal algebra of all vector fields seems to 
ignore any supplementary imposed structure on the manifold.

\section{Other cases}

In the non-uniform linear case, for which \textit{$\kappa $} satisfies condition (3), along with the equations that one derived in the uniform linear case, one has, along with equations (7, 8, 9), the additional equations:
\begin{displaymath}
$L$_{X}$\textit{$\kappa $}$_{\mu \nu }$,$_{\lambda }^{\alpha \beta }$ = \textit{$\gamma $}$_{1}$
\textit{$\kappa $}$_{\mu \nu }$,$_{\lambda }^{\alpha \beta }
\end{displaymath}
\begin{displaymath}
F_{\rho \sigma } (X^{\lambda }$,$_{\alpha }$
\textit{$\kappa $}$_{\mu \nu }$,$_{\gamma }^{\rho \sigma } +
X^{\mu }$,$_{\beta }$
\textit{$\kappa $}$_{\mu \gamma }$,$_{\alpha }^{\rho \sigma } + 
X^{\nu }$,$_{\gamma }$
\textit{$\kappa $}$_{\beta \nu }$,$_{\gamma }^{\rho \sigma}$ ) = 
\textit{$\gamma $}$_{2}$ $F_{\rho \sigma }$
\textit{$\kappa $}$_{\mu \nu }$,$_{\gamma }^{\rho \sigma }
\end{displaymath}
\begin{displaymath}
X_{\mu \nu }$,$_{\lambda }$ = 
\textit{$\alpha $} $F_{\alpha \beta }$
\textit{$\kappa $}$_{\mu \nu }$,$_{\lambda }^{\alpha \beta }
\end{displaymath}
\begin{displaymath}
X_{\alpha \beta }$
\textit{$\kappa $}$_{\mu \nu }$,$_{\lambda }^{\alpha \beta } $ = 
\textit{$\beta  $}$F_{\alpha \beta }$
\textit{$\kappa $}$_{\mu \nu }$,$_{\lambda }^{\alpha \beta }
\end{displaymath}

Due to the complexity of this system, the results are even more inconclusive 
than the previous ones, in advance of more details regarding the nature of 
\textit{$\kappa $}.  This should not be too surprising, since one is essentially 
dealing with the realm of linear wave optics, which is certainly complicated 
in its own right.

In the uniform non-linear case, for which \textit{$\kappa $} satisfies condition (4), the symmetry equations take the form:
\begin{displaymath}
\frac{\partial X^\alpha }{\partial x^\mu }= $ 
\textit{$\alpha $} \textit{$\delta $}$^{\alpha }_{\mu }$ +
\textit{B}$^{\alpha }_{\mu } \quad $
(\textit{B}$^{\alpha }_{\mu} \equiv $
\textit{$\alpha $}$^{\nu }_{\beta }$
\textit{K}$_{\mu \nu }^{\alpha \beta}$,
\textit{$\alpha $}$^{\nu }_{\nu } \ne  0)
\end{displaymath}
\begin{displaymath}
\frac{\partial X_{\mu \nu } }{\partial F_{\alpha \beta } }=$ 
\textit{$\alpha $}$_{2}$
\textit{$\delta $}$^{\alpha }_{\mu }$
\textit{$\delta $}$^{\beta  }_{\nu }$ +
\textit{$\beta  $}$_{1}$
\textit{K}$_{\mu \nu }^{\alpha \beta }
\end{displaymath}
\begin{displaymath}
$L$_{X}$
\textit{K}$_{\mu \beta }^{\alpha \beta }$ =
\textit{$\alpha $}$_{2}$
\textit{$\delta $}$^{\alpha }_{\mu }$
\textit{$\delta $}$^{\beta }_{\nu }$ +
\textit{$\beta  $}$_{2}$
\textit{K}$_{\mu \nu }^{\alpha \beta }
\end{displaymath}

The form of these equations is essentially the same as in the uniform linear 
case, except that now \textit{$\kappa $} has been deformed to $K$ by way of \textit{$\kappa $}$_{\mu \nu }^{\alpha \beta ,\kappa \lambda }$. Furthermore, $K$ is not assumed to have constant components, so solving the system of PDE's is more involved than in the uniform linear case.

\section{Discussion}

Although the results of the foregoing analysis are not unambiguously definitive they are certainly probative.  This is because there is an intimate connection between the symmetries of electromagnetism and the geometry of the spacetime manifold.  If indeed the reduction of these symmetries to the familiar Lorentzian symmetries is associated with a corresponding reduction of the geometry (in the sense of $G$-structures on manifolds) then resolving the issue of the enlarged pre-metric electromagnetic symmetry group is also associated with the question of what spacetime geometry should properly enlarge the scope of the Lorentzian geometry that Einstein established by way of his theory of gravitation.

The strongest indications are that one must simply learn to think of the geometry of the spacetime manifold in terms of the complex projective geometry of the bundle of 2-forms when it is given an almost-complex structure instead of the real metric geometry of the tangent bundle when it is given a Lorentzian structure.  This also involves considering the tangent 2-planes (if not, complex lines) that are described by decomposable 2-forms as being the elementary geometric objects, rather than the tangent vectors that one considers in conventional Riemannian geometry.  Perhaps this also produces an enlargement of the mechanics of matter that moves along congruences of curves to the mechanics of matter that moves along higher-dimensional foliations of wavefronts.

\textbf{Acknowledgements} 

The author would like to thank Academician Anatoly Nikitin and the Institute 
of Mathematics of the National Academy of Sciences of Ukraine for their 
hospitality and their administration of a stimulating and memorable 
inter-disciplinary conference. He would also like to thank George Bluman, B. 
Kent Harrison, Hans Peter Kunzle, Peter Clarkson, Elizabeth Mansfield, and 
Anjan Biswas for illuminating discussions on the symmetries of the 
differential equations of physics, as well as the Burmeister Foundation at 
Bethany College for providing financial assistance with the travel expenses 
associated with attending the conference.

\end{document}